# Ensembles, Dynamics, and Cell Types: Revisiting the Statistical Mechanics Perspective on Cellular Regulation

Stefan Bornholdt [a,*] and Stuart Kauffman [b]

[a] Institute for Theoretical Physics, University of Bremen, 28359 Bremen, Germany
[b] Institute for Systems Biology, Seattle, WA 98109, USA

**Abstract**
Genetic regulatory networks control ontogeny. For fifty years Boolean networks have served as models of such systems, ranging from ensembles of random Boolean networks as models for generic properties of gene regulation to working dynamical models of a growing number of sub-networks of real cells. At the same time, their statistical mechanics has been thoroughly studied. Here we recapitulate their original motivation in the context of current theoretical and empirical research. We discuss ensembles of random Boolean networks whose dynamical attractors model cell types. A sub-ensemble is the critical ensemble. There is now strong evidence that genetic regulatory networks are dynamically critical, and that evolution is exploring the critical sub-ensemble. The generic properties of this sub-ensemble predict essential features of cell differentiation. In particular, the number of attractors in such networks scales as the DNA content raised to the 0.63 power. Data on the number of cell types as a function of the DNA content per cell shows a scaling relationship of 0.88. Thus, the theory correctly predicts a power law relationship between the number of cell types and the DNA contents per cell, and a comparable slope. We discuss these new scaling values and show prospects for new research lines for Boolean networks as a base model for systems biology.

## I. Introduction

We wish to explore a statistical mechanics perspective for biological regulation, in face of the overwhelming fact of natural selection. When Maxwell used statistical mechanics in physics to derive the velocity distribution of gas molecules, he did not have to worry that natural selection might alter that distribution. But as many note, biology without evolution is unthinkable, and evolution is substantially driven by natural selection. Evolution thus explores a large space of possible evolutionary paths, where some of them are realized in nature as existing species, while many possible paths are not.

[*] bornholdt@uni-bremen.de

The part of evolutionary history that we find recorded in the fossil record points to the fundamental role that genetic regulation plays in orchestrating the evolved body plans of species (Valentine 2000). The central machineries of this regulation are networks of genetic and biochemical interactions in the living cell. Their dynamics has been studied ever since the first genetic switch became evident (Jacob and Monod 1961) and continues to our genomic age today.

Early models for such a systems level perspective of interconnected biochemical switches already used the Boolean language and logical gates of early computer science (Sugita 1961, 1963a, 1963b, 1966). How can a circuit of switches provide cells with the homeostasis and stability needed to form and run an organism, yet provide the flexibility and differentiation to account for the wonderful biological diversity in cell types and the mechanisms of morphogenesis? Lacking most of the circuitry of regulation, these were hard questions at that time and in one way or the other are difficult still today. Bottom up studies of specific circuits and dynamics of small networks pointed to the central role of feedback in regulation and provided for a solid basis of mechanisms that account for the main biological phenomena, as beautifully worked out by early works of René Thomas and his subsequent work with collaborators which initiated work on bistable and multistable genetic circuits (Thomas 1973; Thomas et al. 1995; Thieffry and Thomas 1997).

A complementary approach to learn about possible mechanisms of regulation in large networks has been the top-down approach: Simulating large Boolean networks (or: networks of switches) as toy models for potential system's behavior of large regulatory networks. Given the then near total lack of knowledge about the circuitry of biological networks, still ensembles of random Boolean networks could be explored in order to reach hypotheses about real networks and guide their understanding and further exploration (Kauffman 1969). Under what conditions could stability and differentiability coexist in these networks?

Surprisingly, in this study performed by one of us fifty years ago (Kauffman 1969), an unexpected specific property of random Boolean networks turned up which is due to the physics of networks of switches: The interplay of dynamics and geometry in such networks sharply separates their dynamics between order and chaos, depending on the connectivity and the nature of interactions between the genetic switches. This so-called phase transition in the dynamics of a network results in scaling laws that are observable and have practical consequences. A typically moderate number of stable attractors of the dynamics provide a potential mechanism for regulating cell types in a subroutine-like fashion, as in software. The attractor "landscape" found in large random Boolean networks is a natural extension of the multistationarity as characterized in small systems (Thieffry and Thomas 1997). As a new result, random Boolean networks generate testable hypotheses of a specific quantitative scaling between the number of genes in a network and the size and number of attractor states in their dynamics. This came as a big surprise.

What is the status of these findings now, after 50 years?

Here, we wish to re-investigate this specific physics perspective of random Boolean networks in the light of current experimental evidence of genetic regulatory networks, cell types as attractors of the dynamics of such networks, and the generic properties of the ensemble of networks that evolution is exploring. In particular we will revisit the seminal paper: ``Metabolic stability and epigenesis in randomly constructed genetic nets'' by co-author S.A.K. in 1969 who, alongside with the statistical study of random Boolean network dynamics, confirmed the curious statistical scaling relationship between genome size and, both cell cycle length and the number of cell types, by compiling data for a range of organisms. This led to the hypothesis that this biological scaling could have an explanation in the statistical mechanics of an underlying, evolved circuitry.

Since those early days, Boolean networks, or networks of switches, as a template for gene regulation of cell types have been a guiding vision for decades. Now, half a century later, the attractor hypothesis (that cell types are in fact attractors of an underlying genetic network dynamics) can be safely seen as an established fact, the latest triumphant support being the reprogramming of differentiated cells into induced pluripotent stem cells (iPS) and the switching between cell types by specific protein cues (Chang et al. 2011).

What have been the main developments from the perspective of random Boolean networks since? First of all, genomic data provides a much better estimate of overall gene numbers and their roles in the genome. With respect to Boolean network models, however, two particular developments are notable.

One new development is that modeling gene regulation with Boolean networks, today, left its legacy role as a mere anecdotic model of gene regulation. As the circuitries of gene regulatory networks are partly known today, we know certain modules of the regulatory network in sufficient detail for dynamical simulation. For such network modules, the time sequence patterns of regulatory genetic networks from living cells are exactly reproduced by the corresponding Boolean networks, indeed! This is truly surprising as it does not require any kinetic constants, as was first noted by Albert and Othmer (2003b). Today, examples of such models range from cell cycle control networks, developmental circuits, to apoptosis and cancer networks, among others (Mendoza et al. 1999; Sánchez and Thieffry 2001,2003; Albert and Othmer 2003; Espinosa-Soto 2004; Li et al. 2004; Zhang 2008; Davidich and Bornholdt 2008a, 2013; Faure and Thieffry 2009; Krumsiek et al. 2011; Lovrics et al. 2014; Zañudo and Albert 2015; Ríos et al. 2015; Zhou et al. 2016b; Arias Del Angel et al. 2018; Yachie-Kinoshita et al. 2018; Zañudo et al. 2018). Even a tinker toy set for simulating small genetic networks from biology with hardware Boolean networks is available today (Bornholdt and Kopperschmidt 2015). Network dynamics of real genetic regulatory networks being tractable by Boolean networks is an impressive confirmation of the cellular attractor hypothesis.

A second development is that the random Boolean network approach to a first systems biology view on regulatory networks and its findings of scaling laws sparked a physics subfield of "statistical mechanics of random Boolean networks" over the following decades. It is worthwhile taking a look at their results in the biological context which is one motivation for this article. What did we learn since the early random Boolean network models? What can we add to the random network ensemble perspective?

Clearly, selection does not pick random regulatory networks, the networks we find in Nature do not comprise a random ensemble. The random networks ensemble assumption was the simplest best guess given the complete lack of data about the circuitry at that time. Biological networks are always functional networks, and in the examples of known regulatory sub-networks we find "functional ensembles" as Lau, Ganguli, and Tang coined so adequately, when they studied the yeast cell cycle network functional ensemble (Lau et al. 2007). Indeed they found that out of the $5.39 \times 10^{57}$ possible networks in their Boolean network model with 11 nodes, an impressive set of $5.11 \times 10^{34}$ network variants produce the exact dynamical trajectory needed for cell cycle control. Their study of this functional ensemble clearly contributes to an understanding of the evolution of regulatory networks. The question of why Nature chose a particular network out of this ensemble leads to interesting research paths and questions about selection and evolution of cellular regulation.

The ensemble approach thus gains a number of new facets, from random to functional and further, in the presence of biochemical noise, to robust, or reliable, ensembles, a concept and language that allows to sort through the regulatory networks selected – or not selected - by evolution (Klemm and Bornholdt 2005b). This may even explain why Boolean networks reproduce the dynamical trajectories of biological regulatory modules so well (which they normally don't for just random Boolean networks with noise): Natural selection seems to operate in an ensemble of reliable networks optimized to work in the rough environment of biochemical "hardware". That, in return, allows for other ways of sloppy modeling (Transtrum et al. 2015) as, in our case, by Boolean networks.

These new developments in applied and theoretical research on Boolean networks call for a renewed look at the 1969 paper. Key questions are, first, whether the predictions of scaling in attractor numbers and lengths in simple ensembles of just random Boolean networks and the claim of criticality in regulatory networks are still valid in the light of the physics of random Boolean networks, and, second, whether the empirical scaling law for the number of cell types versus DNA per cell that holds across many phyla can be confirmed in the light of new data.

There is now solid evidence that evolution indeed has achieved and is maintaining genetic networks near dynamical criticality (Balleza et al. 2008; Daniels et al. 2018). Is this a curiosity or a biological necessity as early models suggest? Beyond the obvious reason to stay away from a frozen, as well as a chaotic network when processing information, there are a number of further questions stemming from the physics of critical networks. As we shall see, the generic properties of this ensemble include aspects of cell stability to perturbations, i.e. homeostasis, the fact that any cell type can differentiate by flow to another attractor to only a few other cell types, and the fact that there are pathways of differentiation from one cell type, directly or indirectly, to many others. All these features are true of real cells in multicellular organisms.

As criticality seems to persist, it will be interesting to revisit the scaling law for the number of cell types versus DNA per cell that hold across many phyla, a central theme of the 1969 paper. We will discuss it in the light of current empirical data (van Nimwegen 2006) and use recent knowledge about the relationship of DNA mass vs. the number of regulatory genes to make empirical scaling and random Boolean network scaling exactly comparable for the first time.

This paper is organized as follows. In section 2 we ask what a cell type is, and respond that it is an attractor in the dynamics of the genetic regulatory network that controls ontogeny. Multiple cell types requires multiple attractors in the network.

In section 3 we introduce Boolean network models of genetic regulatory networks, define ensembles of such networks, and describe the behaviors of such a network. We recapitulate the results of statistical mechanics of Boolean networks and some applications.

In section 4 we discuss that data now are strongly suggesting that genetic regulatory networks are close to dynamical criticality. We discuss possible selective advantages of being critical.

In section 5 we discuss the generic properties of critical networks with respect to cell type stability, differentiation, and expected dynamical meso-scale properties of critical networks. We will revisit scaling of cell type numbers with DNA per cell and compare the scaling law to current data. Remarkably, the number of model cell types as a function of the number of genes scales as a power law. So too does data on the number of cell types as a function of DNA per cell across many phyla. The critical ensemble predicts the power law slope of the scaling of number of cell types with DNA per cell is 0.63 (including the effects of supralinear scaling of regulatory genes per DNA) not far from the observed scaling of 0.88, supporting the case of an ensemble theory for the networks of cellular regulation.

## II. What is a Cell Type?

In 1960, a central problem in biology asked how it could be that all the cells in a multicelled organism have the same DNA, yet different genes are expressed in different cell types. Red cells make hemoglobin, white cells make antibody molecules. In 1961 two French molecular biologists, Jacob and Monod, solved the problem with the lactose operon (Jacob and Monod 1961). Here a repressor protein, R, binds to a cis site, the operator, of the lactose operon. When the operator is bound, the three adjacent structural genes of the operon cannot be transcribed, so are inactive. However, if lactose enters the cell, lactose binds the repressor protein, pulling it off the operator. Now the adjacent structural genes can be transcribed, including the gene coding for beta galactosidase, the enzyme that metabolizes lactose. Jacob and Monod won a Nobel prize for this work.

In 1963 Jacob and Monod published a seminal paper (Jacob and Monod 1963), pointing out that a hypothetical genetic circuit in which gene A represses the expression of gene B, while B represses the expression of A. This circuit is bistable for it has two alternative steady states of gene activity: A on B off, A off B on. Thus, the same genes could be in two different states of gene expression, solving the mystery of how the same genes in each cell could support different patterns of gene expression.

Two conclusions followed. A on B off and A off B on are two different "attractors" of this little bistable genetic circuit. Alternative cell types must be different attractors of some genetic regulatory network. Second, there must be some large genetic regulatory network, with alternative attractors constituting multistability. On this view, a cell type is an attractor and differentiation is passing from one attractor to another by signals or noise.

Astonishingly, this insight from Jacob and Monod in 1963 has been substantially lost in the enormous advances in molecular biology that is often a kind of molecular anatomy. We think that protein A touches protein B which touches protein L, so have the idea of pathways and interlinked pathways. But, in a way, we may have partially lost the insight from Jacob and Monod that a genetic regulatory network is a dynamical system with, presumably, multistability and multiple attractors. Lacking the idea of attractors we have no clear idea of what a cell type is.

René Thomas initiated seminal work on the conditions required for the emergence of bistability or multistability in small genetic networks in 1981. In general, this requires positive feedback (Thomas 1981). In their review, Gagneur and Cassari (2005) comment, "This hypothesis formulated from the study of simple Boolean models has subsequently led to the discovery of a fundamental principle on feedback that holds true for more general systems of differential equations equations."

A large body of work has now been done on the conditions for multistability as summarized in Gagneur's recent review (Gagneur and Cassari 2005). In the present paper we shall show that critical random Boolean networks not only exhibit multistability, but lead to scaling laws for phenomena such as the relation between the number of cell types in an organism and its DNA per cell (Kauffman 1969).

There is now, however, good evidence that cell types are high dimensional attractors. Huang and collaborators (Huang et al. 2005) took HL60, and induced differentiation to polymorphoneuterophil, PMN, using vitamin A and another substance. They followed gene expression of all 23000 genes using gene arrays at three time points for both treatments. This shows that the gene expression pattern diverged for the two treatments at the temporal midpoint, then converged to the same new expression pattern corresponding to being a PMN. So trajectories converged on the same new pattern of expression from two different directions in high dimensional space, demonstrating that target pattern is an attractor of the dynamics.

A similar experiment was carried out by Zhou et al. (2016a). 1500 FDA approved drugs were screened for any that were able, alone, to induce differentiation of a breast cancer cell line to become adult breast cells and stop dividing: Indeed, 16 of the drugs were able to do so. Importantly, we can get a cancer cell to differentiate into a "normal" non-dividing cell, so cancer differentiation therapy is possible. For each of the 16, single cell gene expression analysis of the early, middle and late time response to the chemical perturbation has been recorded. Like HL60, gene expression patterns diverged across the 16 at the temporal midpoint, but converged to the same new attractor at the final time point.

We conclude that there is good evidence that cell types are high dimensional attractors in the dynamics of the genetic regulatory network.

**III Boolean Networks**

In order to consider statistical properties of networks, a particular simplification of regulation by binary switches led to Boolean networks as minimal models, discrete dynamical networks, for regulatory networks. Today, they represent the simplest form of dynamical representation of biochemical regulatory networks (for a review of a suite of models see Tyson et al. 2019) and have gained a new popularity.

A Boolean network is a discrete dynamical system with N binary variables, here "genes". Each gene receives inputs from K genes chosen among the N. K may vary from gene to gene. The assignment of the K inputs to each gene specifies the "wiring diagram" of the network. Then each gene is assigned a Boolean function on its K inputs, specifying whether that gene will turn on or turn off at the discrete next clocked moment, depending upon the on/off values of its K inputs at the present moment. Time is discrete and all the binary variable update their values synchronously at each clocked moment.

A "state" of the network is any of the 2 to the N choices of on or off for each of the N genes. Thus there are 2 to the N states of the network. If the network is placed in an initial state at time t, then at time t + 1, each gene will assess the values of its K inputs, on or off, and determine whether at the next moment, t+1, that gene should be on or be off. Since this is true for all N genes that update at the same moment, the total network passes from the initial state to some successor state among the 2 to the N states. Over time the system traces a "trajectory" or sequence of states. This is a flow in state space. Because there is a finite number of states when updating the network synchronously, the system must eventually hit some state for a second time. Then, because the system is deterministic, the network will "do the same thing" and trace out a recurrent "state cycle" in state space. In the absence of perturbation, the system will keep traversing the state cycle forever. In general multiple trajectories all terminate on the same state cycle with is, therefore, an attractor, attracting flow along the set of trajectories leading to it. The number of states on the state cycle can be 1, a steady state, or any other number up to 2 to the N. The set of states flowing to one attractor is called its "basin of attraction".

Generically, such a network has multiple attractors, each draining a "basin of attraction" of trajectories that flow to that attractor.

If an attractor is a cell type, then differentiation is flow from one to another attractor induced by noise or signal. The stability of attractors can be studied by perturbing each in all possible minimal ways by transiently flipping a gene from on to off, or off to on. This places the network in a perturbed state with respect to that attractor and the net may homeostatically return to the same attractor, or flow to another attractor.

One may object that Boolean networks are stark simplifications of real biochemical networks. However, the non-linear, often switch-like nature of biochemical reactions and the central idea of a bistable genetic switch of Jacob and Monod sparked independent concepts of information processing circuits and their modeling with simplified automata (Kauffman 1969; Rössler 1972; Thomas 1973; Glass and Kauffman 1973). Boolean network automata assume discrete states (on or off), absence of biochemical noise, and synchronous stepwise updates: all are dramatic simplifications of cellular regulation.

Today, half a century after the 1969 paper that made the bold claim of Boolean networks being a relevant model for gene regulation, we find that Boolean nets indeed seem advantageous models that are widely used for modeling genetic regulatory networks. When and why is the Boolean approximation sensible? Most of all, the Boolean approximation is a well-defined mathematical limit of continuous biochemical equations (Glass 1975; Davidich Bornholdt 2008b) that captures the main properties of the attractors of the dynamics. Boolean networks have become a standard model for studying the statistical properties of attractors and the dynamics of networks.

Ensembles of Boolean Networks and Critical Networks.

The 1969 paper sparked the statistical point of view on the dynamics of genetic networks using random Boolean nets. One can crisply specify different ensembles or classes of Boolean networks. In the initial works (Kauffman 1969, 1993) S.A.K. studied networks with a fixed number K of inputs, for K = 2, and K = N.  K = 2 networks have remarkable properties including that they turn out to be dynamically "critical", as defined shortly.  Among the remarkable properties of K = 2 networks is that the number of states on a state cycle scales as the square root of the number of model genes, N.  So a network with 100,000 model genes will cycle among a tiny 318 states out of 2 to the 100,000. This is intense localization in state space and what S.A.K. has called for years, "Order for free". The number of attractors also scales as square root N. This is striking. A net with 100,000 genes would have on the order of 318 attractors, hence cell types. But humans have on the order of 285 cell types by histological criteria!  We shall emphasize the predicted sublinear scaling of attractor numbers versus DNA diversity, however we also note that the absolute values predicted are biologically very reasonable.

As we will see below, the scaling law for the number of attractors compares well with the current scaling law for the number of cell types across phyla. In a log number of cell type versus log DNA per cell plot, current data indicates a power law, with a slope of 0.88 (Niklas 2014). We will claim below that an ensemble theory such as this is a reasonable account of the data. But first we have to discuss the ensemble of genetic networks that evolution is exploring. That ensemble appears to be critical.

If one varies for N and K, one is sampling a changing ensemble of Boolean networks. It turns out that Boolean networks have three regimes: Ordered, Critical and Chaotic.  There is a two dimensional parameter space, K and P. K is the number of inputs per variable. P is the fraction of "1" values in a Boolean  function. In the K P plane, a one dimensional line is critical, and separates the ordered from the chaotic regime. Thus criticality is rare. For P=0.5, i.e. randomly chosen Boolean functions, K = 2 networks are critical. As K increases, P must increase or decrease from 0.5 to achieve criticality.

A particular subset of Boolean networks are (Boolean) threshold networks which has proven to be particularly suitable for simulating  gene regulatory networks, as threshold functions often come quite close to representing transcriptional regulation.  Here, the inputs at each node are summed up and compared to a threshold to activate the node once a certain number of active inputs are received (thus a subset of Boolean functions).  Ensembles of random threshold networks exhibit the same phenomenon of a phase transition between ordered and chaotic dynamics for a comparable critical K, with an activation threshold $\theta$ representing the second parameter that tunes the criticality transition.

The behavior of chaotic random Boolean networks is radically different from critical networks. First the lengths of state cycles scales exponentially in N. For K =N, the length of state cycles is 2 raised to the N/2, the square root of the number of states. A network with only 200 variables would require 2 raised to the 100 moments to traverse its cycle. At a microsecond per state transition, that would be the lifetime of the universe. This is obviously not biologically plausible. Real networks cannot be strongly chaotic. For K = N, the number of attractors scales as N/e. Perhaps most critically, if a single gene is perturbed, switched transiently to the opposite value, a vast avalanche of "damage" spreads to most of the nodes. Define a gene as damaged if it ever behaves differently that it would have without the perturbation and color it purple. Purple avalanches spread across most of the network.

By contrast critical networks have many small and a few large avalanches, distributed in a power law, i.e. log size of avalanche vs. log number of avalanches. The power law is slope -1.5 in critical networks. This affords critical networks the capacity to control their own behavior locally by small avalanches, and far away by rarer large avalanches. One phrase for criticality is "the edge of chaos", a moniker that is easy to like.

As we will see shortly, the size distribution of avalanches in critical nets predicts brilliantly the size distribution of avalanches of damage due to 1200 single gene deletions in yeast. This is evidence that yeast is critical, or perhaps slightly subcritical. More evidence, below, is accumulating that cells are, indeed, critical.

Critical networks have a meso-scale structure that constitutes predictions that should now be testable. First, a large number of genes in a network fall to a "frozen" state, i.e. frozen fixed off, or fixed on. Call this the "Frozen Component". The frozen component is in the same fixed state on all attractors, hence on all cell types. This component percolates across the network leaving behind one or several "functionally isolated twinkling islands". These islands contain genes that are either turning on and off on an attractor, or in different states on different attractors. Each island is functionally isolated because no signal can pass between islands via the frozen component, once it falls to its frozen state. In general each island has more than one attractor of its own. If there are M islands, each with, say, 2 alternative attractors, then the total attractor set of the entire network is 2 to the M, i.e. each way to chose one attractor from each of the M different islands. But this is a kind of combinatorial epigenetic code: A cell type attractor of the entire network is characterized by a specific combination of choices of attractors, one from each of the M islands. These are clear and powerful predictions that should be testable using gene expression data from a set of many cell types, each at a single cell level.

Let us now revisit the 1969 paper and the history of theoretical considerations on critical K = 2 random Boolean networks.

Apart from popularizing the cellular attractor hypothesis, the 1969 paper initiated more than four decades of research on the statistical mechanics of random Boolean networks. The scaling laws of attractor lengths and numbers, the two central points of the paper, have been under scrutiny in the theoretical physics community. This research laid the foundation of a characterization of order and chaos in the dynamics of these networks, and the particular role of the boundary between these two regions, so-called criticality. In physics, criticality characterizes dramatic changes of matter, freezing of water and the like, while in dynamical networks, criticality coincides with a region between dynamical order and chaos, where it has been argued that information processing finds favorable conditions (Bertschinger and Natschlager 2004).

It initiated an avalanche of theoretical papers about the statistical mechanics of those nets, certainly in part because the statistical mechanics of random Boolean networks turned out to be challenging, both, analytically and numerically. A review of this prolific phase has been given by Aldana, Coppersmith, and Kadanoff (2003). It illustrates that the determination of accepted values for the scaling exponents of the number of attractors with network size in critical random Boolean networks has been difficult and under active discussion at the time of the review, more than three decades after the first hypothesis. Large sample-to-sample fluctuations and the lack of self-averaging resulted in an Odyssey of wandering exponents: The long-time consensus of the attractor number exponent, the original square root of N, persisted throughout the 90ies (Aldana et al. 2003). However, at some point the faster computers plus improved numerical methods first showed an exponent of one in 2001 (Bilke and Sjunnesson 2001), until the analytical tour de force of Troein and Samuelsson (2003) proved a superpolynomial scaling of the number of attractors with N. This was to the horror of many earlier researchers, and destroyed one central prediction of the 1969 Kauffman-paper in the blink of an eye. Drossel wrote in her review (Drossel 2008) about the 1969 paper: "...the biological data and the computer simulation data are both incorrect", referring to the then new data that the number of genes is poorly reflected by the amount of DNA and the new superpolynomial scaling results for synchronous random Boolean networks. The 1969 paper seemed dead and useless for predictions, leaving the common textbook interpretation that Kauffman networks, as critical random Boolean networks are sometimes called, are left as an anecdotic surrogate model for genetic networks, solely useful for pedagogical purposes.

However, this extreme scaling of attractor numbers raised a few eyebrows: How relevant is a superpolynomially increasing number of attractors for biology when the majority of attractor basins seem to be so tiny that they did not show up in numerical experiments for three decades? Furthermore it posed the legitimate question how realistic the synchronously updated, deterministic random Boolean network models are for biological processes where, in fact, noise is omnipresent. In particular, the assumption of noiseless dynamics seemed quite strong, an argument that two studies of random Boolean networks with noise made concrete: Greil and Drossel (2005) regain the original claim or attractor number scaling of the 1969 paper when considering random Boolean networks with asynchronous update. For Boolean threshold networks, a neural network like subset of Boolean networks particularly suited for biology (Rybarsch 2012), Klemm and Bornholdt (2005a) found the same sub-linear scaling when noise is turned on in a network of autonomous nodes: The number of attractors scales as square root N. Most attractor basins in the noiseless case seem to be tiny points, that all fuse with the biggest basins under noise. Thus, revision of the results under noise seem to indicate that the original exponent holds: The poor numerics available in 1969 appears to have detected only the big attractor basins, while missing the exponentially many ones only seen in analytical calculations which are extremely small. And in turn, it is these attractors, only, that appear to survive in the presence of noise and those are the ones that are biologically relevant. As a result, we have the picture that the 1969 calculations do not seem to be as unrealistic as thought for many years and the Boolean network picture is back to being compared with today's experimental view of genetic networks. This is the main motivation for reconsidering the classical paper here.

Therefore, rereading the old paper, our question is: What about the old biological data in there? Are there any new perspectives today? This is what we try to discuss in the remainder of this paper.

## IV  Genetic Regulatory Networks are Critical

Very good evidence is accumulating that the genetic regulatory networks in real cells are critical, underlining a core assumption of the 1969 paper.

There are three lines of evidence. First Serra et al. (2007), and also Ramo et al. (2006), separately analyzed 154 deletion mutants in yeast. Each mutant alters the level of expression of some genes. Consider the number of genes whose expression is altered in a damage avalanche. Both Ramo and Serra found that the distribution was a power law, slope -1.5, indicating criticality. More recently, Villani et al. (2018) have analyzed 1200 single gene deletion mutants in yeast, and found from the avalanche distribution that the yeast cell is slightly subcritical, with a sensitivity $\lambda = 0.89 < 1$ in the slightly ordered regime. This large dataset of 1200 mutants now allows to test the type of statistical predictions of an ensemble theory, suggesting that Yeast seems to be just slightly subcritical.

Nykter et al. examined the flow in gene expression state space for nearby initial states. Criticality shows up as flow that is parallel, it neither diverges (chaos), nor converges (order). The data for macrophage gene expression are exactly critical (Nykter et al. 2008).

Daniels et al. (2018) have examined 67 Boolean net models of real cell networks. Almost all are exactly critical, a few are slightly sub- or supracritical.

This now provides good evidence that genetic regulatory networks across a number of organisms and phyla are critical. More data are needed, but we tentatively conclude that genetic regulatory networks are critical or slightly subcritical.

Criticality, the edge of chaos, seems a good place to be. First, the power law distribution of avalanches of damage allows the cell to be stable, but to correct errors nearby and sometimes far away, without veering into chaos. Mutual information is maximized in critical networks (Rebeiro 2007). Such networks evolve new attractors gracefully under mutation (Torres-Sosa et. al. 2012). Fourth, attractor numbers are tiny in vast state spaces, order for free. Fifth, the number of attractors scales as something like square root N. Thus the system has a modest number of alternative modes of behavior which is controllable, in contrast to vastly many attractors. Sixth, criticality allows a network to optimize the balance between not forgetting its past due to high convergence in state space, but being able to act reliably by not being chaotic.

**V. Critical Ensemble Generic Properties and Predictions**

It is possible to study the stability of critical network attractors to all possible minimal perturbations by flipping a gene transiently on or off. For critical networks the perturbed attractor returns to the same attractor for about 90% of such perturbations – this in fact is homeostasis, and arises for free in the critical ensemble. It is entirely lacking in K=N networks. If the system leaves one attractor for each of the set of all minimal perturbations, it only transitions to a few other attractors. This predicts that a cell type can directly differentiate into only a few other cell types. This prediction is true. Further, a multiplicity of different minimal perturbations take the tested attractor to the same (!) new attractor. This predicts that a multitude of perturbations will induce the same step of differentiation. This is also true. For example, let us remind you of Zhou et al. (2016a) who found 16 out of 1500 FDA approved drugs each one of which induced the same breast cancer cell line to differentiate into adult non proliferating cells. By gene expression analysis, all induced the same differentiation step to a new attractor along diverging the converging trajectories in state space.

But more, each attractor can directly reach only a few other attractors, but from those, others can be reached. This predicts pathways of differentiation true of all multicelled organisms. Pathways are not a logical necessity. The zygote of the sponge could directly differentiate into all its several cell types, but even the sponge has developmental pathways.

All these features are true of the critical ensemble, and true of real cells.

In the paper (Kauffman 1969) the most obvious temporal cycle in cells, the cell cycle versus the DNA per cell, is compared to the prediction of the critical ensemble, where the length of state cycles scales as square root N. Strikingly, across phyla, mean cell cycle time scales as a square root of the DNA per cell. In the days of junk DNA the author S.A.K. came to ignore this prediction. With Encode suggesting that most DNA is functional, we find the prediction intriguing.

Also, the 1969 paper compared the prediction that the number of cell types in an organism should scale as the square root of the DNA per cell. S.A.K.'s data on log number of cell types versus log DNA per cell show a linear slope in a log log plot with a slope of about 0.67. For two reasons S.A.K. came to ignore this prediction. First junk DNA, now restored by Encode to mean that DNA per cell seems a reasonable proxy for the number of variables in the system, e.g. coding and non-coding RNA. Second, as was mentioned above, the paper underestimated the number of attractors in critical networks that are synchronous and was criticized for that. But Greil and Drossel (2005), and, independently, Klemm and Bornholdt (2005a) found that for non-synchronous networks the scaling is again compatible with square root N.

Therefore it is interesting to reconsider the observations in the light of new data. In particular, we can use the number of regulatory genes directly today, dropping the imprecise proxy of the amount of DNA in the cell. This is a central point of this paper. The current data across many phyla for log number of cell types versus log DNA per cell is slightly greater than square root N or 0.5 in a log log plot. The current data suggests a scaling of 0.88 (Niklas 2014). Van Nimwegen (2003) has shown that the number of transcriptional regulatory genes $N_{tr}$ grows faster than the total number of genes $N_{DNA}$, as $N_{tr} \sim N_{DNA}^{1.26}$. In a log log plot the slope is 1.26 ± 0.1. We here argue that the generic properties of the critical ensemble predicts a slope of 0.5 × 1.26 (±0.1) = 0.63, according to $N_{cell\ types} \sim N_{tr}^{0.5} \sim N_{DNA}^{(1.26 \times 0.5)} \sim N_{DNA}^{0.63}$, or up to 0.68 within error bars.

We find three aspects remarkable. First, we indeed observe a power law scaling, i.e. a linear log-log scaling for the log of the numbers of cell types as a function of log DNA per cell across many phyla. This feature is a universal quantity independent of the size of the genome and thus points to a deeper, underlying mechanism. If all cells are critical, the generic properties may be shining though and we see a linear relation in a log log plot of cell types per DNA because that is generic to this ensemble.

Second, we find the exponent of the same approximate size, with the error bar only considering van Nimwegen's statistics of data points, 0.68 versus 0.88 (there are no error bars on the 0.88 slope). We estimate the experimentally derived exponent to being close enough to the theoretical value to warrant closer diligence in future experiments and further statistical analysis of existing databases to narrow down on a more precise value, which eventually decides about the hypothesis that the empirical scaling of the number of cell types with genome size is in fact a scaling of numbers of attractors with the size of the regulatory genetic network across organisms. And that those attractors are the software-subroutines of the different cell types of a genome.

Third, we find it remarkable that the theory predicts not only scaling but very much the right number of cell types. For example, if there are 81,000 functional genes in humans, the theory predicts about the 285 cell types seen not thousands of cell types.

**Discussion**

Good evidence now suggests that genetic regulatory networks across many phyla are in fact critical, or slightly subcritical. Then the theory asserts that the generic properties of critical networks will "shine through", not despite selection, but because selection achieves and sustains the critical subensemble. These properties, adumbrated above, are all visible in the cells of multicelled organisms, from homeostasis to the fact that any cell differentiates directly into only a few others, via a multiplicity of different perturbations. This implies the observed pathways of differentiation. If right, selection does not specifically achieve these features, but lives with and molds them to its further sifting.

In addition, there are at least two scaling laws: cycle time should scale as the square root of the number of genes and in fact, cell cycle time does so scale across phyla. In addition, the number of cell types should scale in a log log plot as 0.63 of the DNA per cell, and a value of 0.88 is observed. Why should there be such power law scaling at all across phyla? Are we to think selection sought and achieved this scaling for some selective reason? It seems highly unlikely. A far better hypothesis to propose is that this scaling is a statistical feature of the critical ensemble, where attractor number grows as the mass of DNA to the 0.63 power.

How could criticality emerge in evolution? Critical networks are a set of measure zero in the space of all possible Boolean networks. How might criticality be achieved then? A recent study is surprising. Torres-Sosa (2012), took chaotic networks and took ordered networks. In both cases, they gently mutated the networks by altering a single connection or bit in a Boolean function. They selected those mutant networks that grew a single new attractor. Over generations this procedure converged on critical networks!

Thus criticality may be reached as a by-product of a certain selection mechanism that serves a specific functional purpose: here, a single new attractor for a simple functional purpose. In biology, complex function is what is selected for, exploring the functional ensemble of networks. Criticality may thus emerge alongside, as a side effect of the functional ensemble or as a useful optimal working point of the networks, rather than selected for directly.

What is the role of criticality, beyond being a mere by-product of selection? Criticality is a dynamical state with distinct properties. Such a state can be an attractor of a dynamical process as well as the result of an evolutionary dynamical process. If such a state at the "edge of chaos" has an evolutionary advantage, one would clearly expect this outcome. But also, if a state "near criticality", as being ordered, yet not disconnected, or simply, not chaotic, is evolutionarily favored, then criticality can be a cheap way to steer an evolutionary process, in order to keep it clear of chaos and close to an ordered regime. This is reminiscent of the mechanism of "self-organized criticality" where a phase transition functions as a convenient tool to tune a dynamical system to an intermediate activity regime.

All in all, it seems very unlikely that selection has 'struggled', to achieve criticality and the scaling relations. A quite sensible hypothesis thus is that selection maintains evolution exploring critical networks, and, held at "criticality" by that selection.

**Conclusion**

We have discussed the 1969 paper by one of us, Kauffman. Given the work on asynchronous networks, the main themes of that paper seem cogent for biology some 50 years later.

We reconsidered the scaling exponent for the number of cell types vs. DNA, where the current experimental value (0.88) increased over the original estimate, while we corrected the model exponent value (from 0.5) to 0.63 by considering a current estimate of the fraction of regulatory genes per DNA.

Why the difference? We do not consider these values as final, yet they point to an interesting direction. The difference in exponents may well find an explanation in the still sparse data and has to be looked at closer from an empirical viewpoint.

But there are also possible arguments from the theory point of view which may point to properties of evolution of biological regulatory networks that are open questions for future research. First of all, regulatory networks in biology are not random networks. Studies of scale-free random Boolean networks already show how changing the Erdös-Renyi random network ensemble to another connection model ensemble changes the dynamics and complicates the scaling exponent issue (Aldana 2003). Distinct topologies selected by evolution thus may alter scalings as a by-product of functional optimization. An empirical hint to another possible cause is the slightly subcritical dynamics observed in some studies (Villani et al. 2018) that may change scaling exponents away from those of precisely critical systems.

Another range of effects may follow from the different ensembles of networks we are considering. We here compare critical ensembles of random networks to evolutionarily selected ensembles which are far from random but instead fulfill further criteria favorable for selection. While the regulatory networks we so far observe in Nature mostly exhibit criticality or near criticality, the selected ensembles certainly differ in various other properties from the critical random network ensemble. The precise statistical properties of functional ensembles, and of ensembles of networks reliable under noise, are mostly open questions at present. Daniels at al. (2018) give a first account of how natural critical networks dramatically differ from random critical networks in their mean connectivity K, but also in their local causal and logical structures. Natural critical networks are, for example, richer in canalizing Boolean functions. How this possibly affects dynamical observables is largely unstudied and may impact scaling laws as well.

There are further lines of future research motivated by the new advent of Boolean networks in biology. As the many working applications of dynamical sequences of real cells reproduced by Boolean networks as mentioned in the introduction show, Boolean networks can serve as a dynamical blueprint of the dynamics implemented in biochemical networks in Nature. The increasing number of examples and the larger networks modeled by Boolean networks opens a new empirical handle on the generic properties of regulatory network attractors.

In closing, we here considered the critical subensemble of genetic regulatory networks. The generic properties of such systems do in fact fit a number of features observed in cells and in cell differentiation in all multicelled organisms. To embrace this, biologists must grow past the power of molecular biology to grant that the cell is a dynamical system, that cell types are attractors, and moreover, that we can attempt to explain substantial features of cell biology via an ensemble approach where the generic properties explain much of cell and developmental biology, without knowing all the myriad details. Such an ensemble approach, a statistical mechanics over ensembles of systems, or networks, could be very useful indeed.

**References**

R. Albert and H.G. Othmer, The topology of the regulatory interactions predicts the expression pattern of the segment polarity genes in Drosophila melanogaster, J. theor. Biol. 223 (2003) 1-18.

R. Albert and H. G. Othmer, "But no kinetic details are needed," SIAM News **36** (10) (2003b).

M. Aldana, Boolean dynamics of networks with scale-free topology, Physica D: Nonlinear Phenomena 185 (2003) 45-66.

M. Aldana, S. N. Coppersmith, and L. Kadanoff, in Boolean Dynamics with Random Coupling, Perspectives and Problems in Nonlinear Science, edited by E. Kaplan, J. E. Marsden, and R. R. Screenivasan, Springer Applied Mathematical Sciences Series, Springer-Verlag, Berlin, 2003.

J.A. Arias Del Angel, A.E. Escalante, L.P. Martínez-Castilla, M. Benítez, Cell-fate determination in Myxococcus xanthus development: Network dynamics and novel predictions, Develop. Growth Differ. 60 (2018) 121–129.

E. Balleza, E.R. Alvarez-Buylla, A. Chaos, S.A. Kauffman, I. Shmulevich, and M. Aldana, Critical dynamics in genetic regulatory networks: examples from four kingdoms, PLoS ONE 3(6):e2456 (2008).

N. Bertschinger and T. Natschlager, Real-time computation at the edge of chaos in recurrent neural networks. Neural Computation, 16 (2004) 1413.

S. Bilke and F. Sjunnesson, Stability of the Kauffman model, Phys. Rev. E 65 (2001) 016129.

S. Bornholdt and G. Kopperschmidt, Gene Regulation: Experiments with the LECTRON Experimental kit (2015) ISBN 978-3-00-051441-8.

R. Chang, R. Shoemaker, W. Wang, Systematic Search for Recipes to Generate Induced Pluripotent Stem Cells, PLoS Comput. Biol. 7 (2011) e1002300.

B.C. Daniels, H. Kim, D. Moore, S. Zhou, H.B. Smith, B. Karas, S.A. Kauffman, S.I. Walker, Criticality Distinguishes the Ensemble of Biological Regulatory Networks, Phys. Rev. Lett. 121 (2018) 138102.

M.I. Davidich and S. Bornholdt, Boolean network model predicts cell cycle sequence of fission yeast, PLoS ONE 3 (2008a) e1672.

M.I. Davidich and S. Bornholdt, The transition from differential equations to Boolean networks: A case study in simplifying a regulatory network model, J. Theor. Biol. 255 (2008b) 269-277.

M.I. Davidich and S. Bornholdt, (2013) Boolean Network Model Predicts Knockout Mutant Phenotypes of Fission Yeast. PLoS ONE 8 (2013) e71786.

B. Drossel, Random Boolean Networks, in: H.G. Schuster (ed.), Reviews of Nonlinear Dynamics and Complexity 1 (2008) 69-110 (Wiley-VCH).

C. Espinosa-Soto., P. Padilla-Longoria, E.R. Alvarez-Buylla, A gene regulatory network model for cell-fate determination during Arabidopsis thaliana flower development that is robust and recovers experimental gene expression profiles. Plant Cell. 16 (2004) 2923–2939.

A. Faure and D. Thieffry, Logical modelling of cell cycle control in eukaryotes: a comparative study, Mol. BioSyst., 5 (2009) 1569–1581.

J. Gagneur and G. Cassari, From molecular networks to qualitaitative cell behavior, FEBS 579 (2005) 1867-1871.

L. Glass and S.A. Kauffman, The Logical Analysis of Continuous, Non-linear Biochemical Control Networks, .I. theor. Biol. 39 (1973) 103-129.

L. Glass, Classification of Biological Networks by their Qualitative Dynamics, J. theor. Biol. 54 (1975) 85-107.

F. Greil and B. Drossel, Dynamics of Critical Kauffman Networks under Asynchronous Stochastic Update, Phys. Rev. Lett 95 (2005) 048701.

F. Greil, B. Drossel and J. Sattler, Critical Kauffman networks under deterministic asynchronous update, New Journal of Physics 9 (2007) 373.

S. Huang, G. Eichler, Y. Bar-Yam, and D.E. Ingber, Cell fates as high-dimensional attractor states of a complex gene regulatory network, Phys. Rev. Lett., 94 (2005) 128701.

F. Jacob and J. Monod, Genetic regulatory mechanisms in the synthesis of proteins, J. Mol. Biol. 3 (1961) 318-356.

F. Jacob and J. Monod, Genetic repression, allosteric inhibition, and cellular differentiation. In: Locke, M. (Ed.), Cytodifferentiation and Macromolecular Synthesis, pp. 30-64. Academic Press, New York (1963).

S.A. Kauffman, Metabolic stability and epigenesis in randomly constructed genetic nets, Journal of Theoretical Biology 22 (1969) 437-467.

S.A. Kauffman, The Origins of Order: Self-organization and Selection in Evolution. Oxford University Press (1993).

K. Klemm and S. Bornholdt, Stable and unstable attractors in Boolean networks, Phys. Rev. E 72 (2005a) 055101R.

K. Klemm and S. Bornholdt, Topology of biological networks and reliability of information processing, Proc. Natl. Acad. Sci. USA 102 (2005b) 18414-18419.

J. Krumsiek, C. Marr, T. Schroeder, F.J. Theis, Hierarchical differentiation of myeloid progenitors is encoded in the transcription factor network. PLoS ONE 6 (2011) 0022649.

K.-Y. Lau, S. Ganguli, and C. Tang, Function constrains network architecture and dynamics: A case study on the yeast cell cycle Boolean network, Phys. Rev. E 75 (2007) 051907.

F. Li, T. Long, Y. Lu, Q. Ouyang, and C. Tang, The yeast cell-cycle network is robustly designed. PNAS 101 (2004) 4781–4786.

A. Lovrics, Y. Gao, B. Juhasz, I. Bock, H.M. Byrne, et al., Boolean Modelling Reveals New Regulatory Connections between Transcription Factors Orchestrating the Development of the Ventral Spinal Cord, PLoS ONE 9 (2014) e111430.

L. Mendoza, D. Thieffry, and E.R. Alvarez-Buylla, Genetic control of flower morphogenesis in Arabidopsis thaliana: a logical analysis, Bioinformatics 15 (1999) 593-606.

K.J. Niklas, E.D. Cobb, A.K. Dunker, The number of cell types, information content, and the evolution of complex multicellularity, Acta Soc. Bot. Pol. 83(4):337–347 (2014).

E. van Nimwegen, Scaling laws in the functional content of genomes, Trends in Genetics 19 (2003) 479.

E. van Nimwegen, Scaling Laws in the Functional Content of Genomes: Fundamental constants in Evolution? In: Power Laws, Scale-Free Networks and Genome Biology. Molecular Biology Intelligence Unit. Springer, Boston, MA (2006).

N.M. Nykter, N.D. Price, M. Aldana, S.A. Ramsey, S.A. Kauffman, L.E. Hood, O. Yli-Harja, and I. Shmulevich, Gene expression dynamics in the macrophage exhibit criticality, PNAS 105 (2008) 1897– 1900.

P. Ramö, J. Kesseli, and O. Yli-Harja, Perturbation avalanches and criticality in gene regulatory networks, Journal of Theoretical Biology, 242:164-170 (2006).

A. Ribeiro, S.A. Kauffman, J. Lloyd-Price, B. Samuelsson, and J.E. Socolar, Mutual Information in Random Boolean Models of Regulatory Networks, Phys. Rev. E 77 (2007) 011901.

O. Ríos, S. Frias, A. Rodríguez, S. Kofman, H. Merchant, L. Torres and L. Mendoza, A Boolean network model of human gonadal sex determination, Theoretical Biology and Medical Modelling 12 (2015) 26.

O.E. Rössler, Basic Circuits of Fluid Automata and Relaxation Systems, Zeitschrift für Naturforschung 27b (1972) 333.

M. Rybarsch and S. Bornholdt Binary threshold networks as a natural null model for biological networks, Phys. Rev. E 86, (2012) 026114.

B. Samuelsson and C. Troein, Superpolynomial Growth in the Number of Attractors in Kauffman Networks, Phys. Rev. Lett. 90 (2003) 098701.

L. Sánchez and D. Thieffry, A logical analysis of the Drosophila gap-gene system. J. Theor. Biol. 211 (2001) 115–141.

L. Sánchez and D. Thieffry, Segmenting the fly embryo: a logical analysis of the pair-rule cross-regulatory module. J. Theoret. Biol., 224 (2003) 517–537.

R. Serra, M. Villani, A. Graudenzi, and S. A. Kauffman, Why a simple model of genetic regulatory networks describes the distribution of avalanches in gene expression data, Journal of Theoretical Biology 246:449–460 (2007).

M. Sugita, Functional analysis of chemical systems in vivo using a logical circuit equivalent, Journal of Theoretical Biology 1 (1961) 415-430.

M. Sugita, Functional analysis of chemical systems in vivo using a logical circuit equivalent. II. The idea of a molecular automaton, Journal of Theoretical Biology 4 (1963) 179-192.

M. Sugita, Functional analysis of chemical systems in vivo using a logical circuit equivalent. III. Analysis using a digital circuit combined with an analogue computer, Journal of Theoretical Biology 5 (1963) 412-418.

M. Sugita, Functional analysis of chemical systems in vivo using a logical circuit equivalent IV. Simulation of cellular control systems using a hybrid computing system, Journal of Theoretical Biology 13 (1966) 330-356.

D. Thieffry, R. Thomas, Qualitative analysis of gene networks. In R.B. Altman et al. (Eds.), Third Pacific Symposium on Biocomputing '98: Maui, Hawaii, USA (pp. 77-88). Singapore: World Scientific (1997).

R. Thomas, Boolean Formalization of Genetic Control Circuits, J. theor. Biol. 42 (1973) 563-585.

R. Thomas, On the relation between the logical structure of systems and their ability to generate multiple steady states or sustained oscillations. Springer Ser. Synergetics 9 (1981)180-193.

R Thomas, D Thieffry, M Kaufman, Dynamical behaviour of biological regulatory networks—I. Biological role of feedback loops and practical use of the concept of the loop-characteristic state, Bulletin of Mathematical Biology 57 (1995) 247-276.

C. Torres-Sosa, S. Huang, and M. Aldana, Criticality Is an Emergent Property of Genetic Networks that Exhibit Evolvability. PLoS Comput Biol 8 (2012) e1002669.

J.W. Valentine, Two genomic paths to the evolution of complexity in bodyplans, Paleobiology, 26 (2000) 513-519.

M.K. Transtrum, B.B. Machta, K.S. Brown, B.C. Daniels, C.R. Myers, and J.P. Sethna, Perspective: Sloppiness and emergent theories in physics, biology, and beyond, The Journal of Chemical Physics 143 (2015) 010901.

J. J. Tyson, T. Laomettachit, P. Kraikivski, Modeling the dynamic behavior of biochemical regulatory networks, Journal of Theoretical Biology 462 (2019) 514–527.

Villani, M., La Rocca, L, Kauffman, S, Serra R. (2018). Dynamical criticality in gene regulatory networks, Complexity 2018, 5980636.

Yachie‐Kinoshita, K. Onishi, J. Ostblom, M.A. Langley, E. Posfai, J. Rossant, P.W. Zandstra, Modeling signaling‐dependent pluripotency with Boolean logic to predict cell fate transitions, Mol. Syst. Biol. 14 (2018) e7952.

J.G.T. Zañudo and R. Albert, Cell fate reprogramming by control of intracellular network dynamics, PLoS Comput. Biol. 11 (2015) e1004193.

J.G.T. Zañudo and S.N. Steinway, and R. Albert, Discrete dynamic network modeling of oncogenic signaling: Mechanistic insights for personalized treatment of cancer, Current Opinion in Systems Biology 9 (2018) 1–10.

R. Zhang, M.V. Shah, J. Yang, S.B. Nyland, X. Liu, J.K. Yun, et al.: Network model of survival signaling in large granular lymphocyte leukemia. Proc. Natl. Acad. Sci. USA 105 (2008) 16308–16313.

J.X. Zhou, Z. Isik, C. Xiao, I. Rubin, S.A. Kauffman, M. Schroeder, and S. Huang, Systematic drug perturbations on cancer cells reveal diverse exit paths from proliferative state, Oncotarget 7 (2016a) 7415–7425.

J.X. Zhou, A. Samal, A. Fouquier d'Hérouël, N.D. Price, S. Huang, Relative stability of network states in Boolean network models of gene regulation in development, BioSystems 142-143 (2016b) 15–24.